\documentclass[12pt]{article}
\usepackage{amsmath,amssymb,graphicx}
\usepackage{amsthm}

\newtheorem{lem}{Lemma}
\newtheorem{thm}{Theorem}

\newcommand{\pr}{\noindent{\bf Proof}. }
\newcommand{\rem}{\noindent{\bf Remark}. }

\newcommand{\pa}{\partial}

\newcommand{\const}{\textrm{const}}

\newcommand{\supp}{ \mathrm{ supp  }}
\newcommand{\hs}{ \hspace{1cm}}

\newcommand{\B}{\Big}

\newcommand{\be}{\begin{equation}}
\newcommand{\ee}{\end{equation}}
\newcommand{\bs}{\begin{split}}
\newcommand{\es}{\end{split}}

\newcommand{\bx}{\mathbf{x}} 
\newcommand{\by}{\mathbf{y }} 
\newcommand{\bk}{\mathbf{k}}

\newcommand{\al}{\alpha}
\newcommand{\De}{\Delta}
\newcommand{\de}{\delta}

\newcommand{\Ga}{\Gamma}

\newcommand{\Om}{\Omega}

\newcommand{\cC}{{\cal C}}
\newcommand{\cD}{{\cal D}}

\newcommand{\cO}{{\cal O}}

\newcommand{\cH}{{\cal H}}

\newcommand{\cK}{{\cal K}}

\newcommand{\cL}{{\cal L}}

\newcommand{\cY}{{\cal Y}}

\newcommand{\bbR}{{\mathbb{R}}}
\newcommand{\bbZ}{{\mathbb{Z}}}
\newcommand{\bbC}{{\mathbb{C}}}

\begin{document}

\title{Scattering on  the Dirac  magnetic monopole}
\author{ 
J. Dimock
\thanks{dimock@buffalo.edu}\\
Dept. of Mathematics \\
SUNY at Buffalo \\
Buffalo, NY 14260 }
\maketitle

\begin{abstract}
We construct wave operators and a scattering operator  for the scattering of  a charged particle on the   Dirac  magnetic monopole.
The analysis features a two Hilbert space approach in which the identification operator matches states of the same angular momentum. 
   \end{abstract}


\section{Introduction}

We study the scattering  of a single charged particle (an electron) by  a magnetic monopole.
The magnetic field of the monopole is described by a connection on a $U(1)$ vector bundle  $E$ over $M = \bbR^3- \{0\}$, the electron
wave functions  are sections of that vector bundle in $L^2(E)$, and there is a self-adjoint Hamiltonian $H$ on this space generating the dynamics.     For the scattering problem we  identify asymptotic states 
which are  elements of  $L^2(\bbR^3)$ with dynamics generated by  the free Hamiltonian $ H_0$.   Scattering is described by  M\o ller wave operators which are proved to exist in a two Hilbert space formulation.   They have
the form $ \Om_{\pm} \Psi    = \lim_{t \to \pm \infty}  e^{iHt} J e^{-i H_0t}  \Psi $ where $J: L^2(\bbR^3) \to L^2(E)$ identifies states of the same angular momentum.   We also    include results in which the monople  Hamiltonian $H$  is perturbed by a potential  $V$ yielding a new Hamiltonian $H+V$.

 This  scattering problem has been previously 
treated by Petry  \cite{Pet80},  who takes  asymptotic   states which are  also   sections of $L^2(E)$ rather than $L^3(\bbR^3)$.  
 Petry's choice of the asymptotic dynamics is somewhat ad hoc and difficult to connect 
with the usual free dynamics.    This means his scattering  results are quoted in terms of "scattering into cones", rather  than  the more standard 
asymptotic wave functions.  
Our two Hilbert space  treatment puts  the problem more in the mainstream of scattering theory.  It  opens the door for further developments  such as
the perturbing potential,    which  seems awkward in the Petry  formulation. 

The paper is organized as follows.     In section 2 we review the description of the monopole as a connection on a vector bundle.
In section 3 we review the definition of the free Hamiltonian $H_0$  in spherical coordinates.   In section 4 we give a detailed definition of 
the monopole Hamiltonian  $H$   also in spherical coordinates.    It is  a map on smooth  sections of the vector bundle and we  show that it defines a self-adjoint operator on $L^2(E)$. In section 5  we find continuum eigenfunction expansions for the radial parts of both $H_0$ and $H$, and in section 6 this is used to give 
detailed estimates on the free dynamics.   In section 7  we prove the main result which is the existence of the wave operators.   Finally
in section 8  we extend the scattering result by  including a potential.

\section{The monopole} 

   In spherical coordinates $(r, \theta, \phi)$  the magnetic field
for a monopole of strength $n \in \bbZ$, $n \neq 0$,  is the two-form
($\star$ is the Hodge star operation)  
\be
B =  \star \frac{n}{r^2} dr  =   n \sin \theta d \theta d \phi.     
\ee 
This is singular at the origin but otherwise is closed  ($dB=0$)  as required by Maxwell's equations.    However it is not exact      ($B  \neq dA$ for any $A$).
If it were exact the integral over the unit sphere  $|x|=1$ would be zero,   but $  \int_{|x|=1} B = 4 \pi n$.  Locally  one can take 
\be  \label{acorn1} 
A = - n  \cos \theta   d \phi  
\ee
since then $dA = B$. 
But this is singular at $x_1=x_2 =0$ as one can see from the representation in Cartesian coordinates 
\be \label{acorn2} 
A =  n  \frac{ x_3}{|x| } \frac {x_2 dx_1- x_1 dx_2}{x_1^2 + x_2^2}.
\ee
This is a problem since we need the magnetic potential $A$ to formulate  the quantum mechanics. 

The remedy  is  to introduce the vector bundle $E$ defined as follows.  First it is a manifold and there is a smooth map   $\pi : E \to M $ to $M = \bbR^3- \{ 0\}$ 
such that each   fibre $E_x = \pi^{-1} x $ is a vector space  isomorphic to $\bbC$.  Further  let $U_{\pm}  $ be 
an open covering of  $M$ defined for $0 < \al  <  \frac12 \pi$ as follows.
First in  spherical coordinates   and then 
 in Cartesian coordinates
 \be  \label{four} 
 \begin{split}
U_+ = & \B \{ x \in M: 0 \leq  \theta  <  \frac{\pi}{2} +  \al \B\}   =  \B \{ x \in M:  1 \geq   \frac{x_3}{|x| }  > \cos\B( \frac{ \pi}{2} +  \al \B)  \B\}    \\
U_- = & \B \{ x \in M:  \frac {\pi}{2} -  \al <  \theta    \leq \pi \B\}  = \B \{ x \in M:  \cos \B( \frac{\pi}{2}  -  \al \B)  >  \frac{x_3}{|x| }   \geq  -1 \B\}. \\
\end{split}
\ee
We require that in   each region there is a trivialization (diffeomorphism) 
\be
h_{\pm} :  \pi^{-1}( U_{\pm} )  \to  U_{\pm}   \times \bbC
\ee
such that for $x \in U_{\pm}$ the map  $ h_{\pm} : E_x \to \{ x \} \times \bbC$ is a linear isomorphism.  They are related by
 the transition function  in $U_+ \cap U_-$  
\be \label{piquant}
 h_+ h _ -  ^{-1}   =  e^{ 2in  \phi } 
\ee
where $e^{ 2in  \phi } $
acts on the second entry $\bbC$.  With  the transition functions specified,  
 $E$ can be constructed as equivalence classes in $M \times \bbC$ with $(x,v) \sim (x, e^{2in \phi}v )$  if $x \in U_+\cap U_-$. 

The connection is     defined by a one-forms  $A^{\pm} $  on $U_{\pm} $.   To compensate  (\ref{piquant})  they  are related in $U_ + \cap U_ -$ by the gauge transformation
\be
A^{+ } = A^{-} +  2n d  \phi. 
\ee
This is accomplished by  taking instead of (\ref{acorn1}),(\ref{acorn2})  
\be   \label{walnut} 
A^{\pm}  = - n( \cos \theta \mp 1 ) d \phi   =  n \B( \frac{ x_3}{|x| } \mp 1 \B) \frac {x_2 dx_1- x_1 dx_2}{x_1^2 + x_2^2}.   
\ee
Each of these satisfy  $d A^{\pm}  = B$,  but now they have no singularity.     Indeed on $U_+$ we have  for points with $x_3 >0$
\be  \label{bingo}
\B| \frac{ x_3}{|x| }    - 1 \B| \leq  \cO \B( \frac  {x_1^2 + x_2^2} { x^2_3}\B). 
\ee
So for fixed $x_3>0$ there is no singularity at $x_1=x_2=0$.   Points in $U_{+}$  with $x_3 \leq 0$  also have $x_1^2 + x_2^2 >0$ so the
singularity is avoided.   Similarly $A^-$ has no singularity on $U_-$.

Now we  can define  a covariant derivative on sections of $E$.  A section of $E$ is a map $\psi : M \to E$ such that $\pi (\psi (x) )= x$.  The set of all smooth sections is
denoted $\Ga ( E)$.   For  $f  \in \Ga(E)$   we define   $\nabla_k  f \in  \Ga(E) $
by specifying that  for $x \in U_{\pm} $ if $h_{\pm} f(x)  = (x,f_{\pm}(x)) $ then  $\nabla_k f$ satisfies  $h_{\pm} (\nabla_k f (x) ) = (x,  (\nabla_k f )_{\pm}(x) )$ where
\be \label{sour} 
  (\nabla_k f ) _{\pm} =   ( \pa_k - i A_k^{\pm} ) f_{\pm}. 
\ee   
Here $A_k^{\pm} $ are the components in $A^{\pm}  = \sum_k A^{\pm} _k dx_k$.  
This defines a section since  in $U_+ \cap U_-$ we have $A_k^{+ } = A_k^{-} +  2n \  \pa \phi/ \pa x_k$  so  
\be
( \pa_k - i A_k^{+} ) e^{2in \phi}  =  e^{2in \phi}  ( \pa_k - i A_k^{- } ).  
\ee
Thus   if  $ f$ is a section,  then $f_+ =   e^{2in \phi} f_-$,  then  $(\nabla_k f )_+    =  e^{2in \phi}   ( \nabla_k f )_-  $, and  hence the pair $(\nabla_k f )_{\pm}$
define a section.

\section{Free Hamiltonian}

We first review the standard treatment of the free Hamiltonian.  This will recall some facts we need and  provide a model for the treatment of the  monopole Hamiltonian.
The free Hamiltonian on $L^2(\bbR^3)$ is  minus  the Laplacian:
 \be 
 H_0 = - \De = - \sum_i \pa_i \pa_i 
\ee
 defined  initially  on  smooth functions.

We study it as a quadratic form and begin  by   breaking  it into radial and angular parts by    
\be \label{sync} 
\begin{split}
( f,H_0 f )  = &  \sum_i \| \pa_i f  \|^2 \\
= &\sum_{i,j}  (\pa_i f  ,   \frac{  x_ix_j}{ |x|^2 } \pa_jf )  +  \sum_{i,j}  \B( \pa_if,   ( \de_{i j} - \frac{  x_i x_j}{ |x|^{2} }) \pa_j f \B) \\
  = &  \|  \frac{1}{|x|}(x \cdot \pa ) f \|^2   +   \sum_i  \| \frac{1}{|x|}   (x \times \pa)_i        f \|^2. \\
   \end{split}
\ee
The last   step follows from  $(x \times \pa)_i =\sum_{jk}  e_{ijk} x_j\pa_k$ and $\sum_i  e_{ijk} e_{i \ell m} = \de_{j \ell} \de_{km} - \de_{j m} \de_{k \ell} $. 
($e_{ijk} $ is the Levi-Civita symbol.)
The skew-symmetric  operators $  (x \times \pa)_i   $  are recognized as a basis for the representation
of the Lie algebra of the rotation group $SO(3)$ generated by the action of the group on $\bbR^3$. 
In quantum mechanics the symmetric   operators    $L_i = -i  (x \times \pa)_i $ 
are identified as the angular momentum.   They satisfy the commutation relations   $[L_i,L_j] = \sum_k  e_{ijk} i L_k$   or   $[L_1,L_2] = i L_3$, etc. 
Now we have
\be
( f,H_0 f )  
    =   \|  \frac{1}{|x|}(x \cdot \pa ) f \|^2   +   \sum_i  \| \frac{1}{|x|}  L_i       f \|^2.
\ee

Next we change to spherical coordinates.  The $ |x|^{-1}  (x \cdot  \pa f ) $ becomes $\pa f/\pa r$ and the $L_i$
become 
\be
\begin{split}
L_1  =&   i  \B( \sin \phi  \frac{\pa}{ \pa \theta } + \cot \theta \cos \phi \frac{\pa}{ \pa \phi  }  \B) \\
L_2  =&   i  \B( - \cos \phi  \frac{\pa}{ \pa \theta } + \cot \theta \sin  \phi \frac{\pa}{ \pa \phi  }  \B) \\
L_3 = &  - i  \frac{\pa}{ \pa \phi  }. 
\end{split}
\ee
The Hamiltonian  in spherical coordinates, still called $H_0$, has become 
\be
( f,H_0 f)   = \|  \frac{\pa  f}{\pa r}  \|^2   +   \sum_i  \| \frac{1}{r}  L_i   f \|^2.
\ee
The norms are now  in 
the space
\be 
\cH_0  = L^2(\bbR^+ \times  S^2, r^2  dr d \Om   )  =    L^2(\bbR^+ , r^2 dr )  \otimes L^2(  S^2,  d \Om  )  
\ee
where $\bbR^+ = (0, \infty)$ and  $d \Om= \sin \theta d \theta d \phi$ is the  Haar measure on $S^2$.  The $L_i$ are symmetric in $ L^2(  S^2,  d \Om  )  $
and after an integration by parts in the radial variable  we have 
\be
H_0 f=-  \frac{1}{r^2} \frac{\pa} {\pa r} r^2  \frac{\pa f} {\pa r} +  \frac{L^2}{r^2}. 
\ee
Here   $L^2=  L_1^2 + L_2^2 + L^2_3 $    is the Casimir operator for the representation of the Lie algebra of the rotation group on  $S^2$ .  This
is a case where it is equal to minus  the Laplacian on $S^2$
\be
L^2  = - \De_2   = -\B(  \frac {1}{\sin \theta} \frac{ \pa} { \pa \theta  } \sin \theta  \frac{ \pa} { \pa \theta  }   + \frac {1}{\sin^2 \theta}   \frac{ \pa} { \pa \phi  } \B) 
\ee
as can be checked directly.

The spectrum of   $L^2$  on $L^2(S^2, d \Om) $ is studied by considering the joint spectrum of the commuting operators  $L^2, L_3$. 
This is a standard problem in quantum mechanics.  It is also the problem of breaking down the representation of the rotation group
into irreducible pieces.  Just from the commutation relations one finds the $L^2$ can only have the  eigenvalues $\ell(\ell+1)$ with $\ell = 0,1,2, \dots$
and that $L_3$ can only have integer eigenvalues  $m$ with $|m| \leq \ell$.   The  corresponding  normalized  eigenfunctions are the 
spherical harmonics $Y_{\ell, m} (\theta, \phi) $ and  are explicitly constructed in terms of the Legendre polynomials. 
They satisfy
\be
\begin{split} 
L^2 Y_{\ell,m}= &  \ell(\ell+1)   Y_{\ell,m}   \hs   \ell \geq  0         \\
 L_3   Y_{\ell,m} = &  m    Y_{\ell,m}  \hs  \hs    |m| \leq \ell.       \\
  \end{split} 
 \ee
The spherical harmonics are complete so this gives the full spectrum of $L^2,L_3$ and yields a definition of corresponding self-adjoint operators.

Let  $\cK _{0,\ell} $ be the $2\ell+1$ dimensional  eigenspace for the   eigenvalue $\ell( \ell+1)$  of $L^2$.  Then $\cK _{0,\ell} $ is  spanned by  $\{  Y_{\ell, m} \}_{ |m| \leq \ell}   $.    
Then  we can write the Hilbert space as
\be
 \cH_0  = \bigoplus_{\ell =0}^{\infty} L^2(\bbR^+, r^2dr) \otimes \cK_{0, \ell} 
 \ee
 and  on smooth functions in this space    $H_0 = \bigoplus_{\ell}\ ( h_{0, \ell} \otimes I ) $  where 
 \be
h_{0, \ell}  =   - \frac{d^2}{dr^2} - \frac{2}{r} \frac{d}{dr}   +   \frac{\ell(\ell+1) }{r^2}.   
\ee
 We study the operator  $h_{0, \ell} $ further in sections \ref{tick}, \ref{tock}.

\section{Monopole Hamiltonian} 

The Hamiltonian for our problem  is initially defined on smooth sections $f \in \Ga (E)$ 
by
\be
 H f = - \sum_{k=1}^3  \nabla_k \nabla_k   f.
\ee
We want to define it as a self-adjoint operator in $L^2(E)$.   In this section we reduce it to a radial problem as for $H_0$.  The treatment 
more or less follows  Wu and Yang
\cite{WuYa76}.

The Hilbert space $L^2(E)$ is 
defined as follows.    If
$x \in U_{\pm} $ and $v \in E_x$ then  $h_{\pm} v = (x, v_{\pm}) $ and we define 
$|v| = |v_{\pm}|$.  This is unambiguous since if  $x \in U_+ \cap U_-$ then $v_{\pm}$ only differ by
a phase and so  $|v_ +| =|v_-|$.  Similarly if $v,w  \in E_x$ we can define $\bar v w \in \bbC$.
The Hilbert space $L^2(E) $ is all measurable sections $f$ such that the norm  $\|f \|^2 = \int  |f(x) |^2 dx$ is finite with
$(g,f)  = \int \overline{g(x)}f(x) dx$.

The covariant derivative $\nabla_k$  is skew-symmetric in this Hilbert space hence  the Hamiltonian is symmetric.   Indeed if   $\supp f,   \supp g \subset U_{\pm} $
and $h_{\pm} f(x)  = (x, f_{\pm} (x))$, etc. then 
\be  (g, \nabla_k  f) = \int  \overline{  g_{\pm} }  ( \pa_k -i A^{\pm}_k ) f_{\pm}  
= -  \int   \overline {    ( \pa_k -i A^{\pm}_k )  g_{\pm} }f_{\pm}  
= - (\nabla_k g, f). 
\ee  
In the general case we write a section $f$ as a sum $f (x) =  f_+(x)  + f_-(x)$ with $\supp f_{\pm} \subset U_{\pm}$.

Now   write the Hamiltonian as a quadratic form, and
as in (\ref{sync})   break  it  into a radial and angular parts     
\be \label{sync2} 
 ( f,Hf) = \sum_k \| \nabla_k f \|^2  
  =   \|  \frac{1}{|x|}(x \cdot \nabla ) f \|^2   +   \sum_k  \| \frac{1}{|x|}   (x \times \nabla)_k        f \|^2.
\ee

Now    there is a problem.  The operators $(x \times \nabla)_k  $,  although they have something to do with rotations, 
no longer give a representation of the Lie algebra of the rotation group.
The commutators now involve extra terms  $ [\nabla _j,\nabla_k] =- iF_{jk}$ where   $F_{jk}=  \pa_jA^{\pm}_k- \pa_kA^{\pm} _j $  is the magnetic field.  
This is not special to the monopole but occurs whenever there is an external magnetic field. 
The resolution  due to Fierz \cite{Frz44}   is to add a term proportional to the field strength.    
Instead of $- i (x \times \nabla)_k = (x \times -i  \nabla)_k$ 
we define angular momentum operators by 
\be \label{Fierz}
\cL_k =  (x \times -i\nabla)_k  -n\frac{x_k }{|x|}. 
\ee
These are symmetric and  do satisfy the commutators $[\cL_i, \cL_j ]  = i\sum_k e_{ijk} \cL_k $.   This follows from commutators 
like
\be
\begin{split}
[\cL_i, x_j ] = &\ i \sum_k  e_{ijk} x_k \\
[ \cL _i,  \nabla_j ] = &\  i \sum_k  e_{ijk}  \nabla_k.\\
\end{split}
\ee
To give the idea we show that  $ [ \cL _1,   \nabla_2] = i \nabla_3$.  We have 
\be   \label{stun}
 \begin{split} 
 [ \cL _1,   \nabla_2]  = &-i [  ( x \times \nabla)_1 ,   \nabla_2]    -  n   [x_1|x|^{-1} ,   \nabla_2   ]   \\
 = & -i [ x _2 \nabla_3- x_3 \nabla_2  ,   \nabla_2]   - n x_1x_2 |x|^{-3} \\
  = &i\nabla_3  -i x_2  [  \nabla_3,   \nabla_2]   - n x_1x_2 |x|^{-3} \\
  = &i \nabla_3  -  x_2  F_{32}    -  n x_1x_2 |x|^{-3}. \\
\end{split}
\ee
But in $U_{\pm} $ we have   $A_3^{\pm} =0$ and taking $A_2^{\pm} $ from  (\ref{walnut})
\be
\begin{split} 
x_2F_{32} = &x_2 \pa_3 A^{\pm} _2 \\
= & x_2  n\pa_3  \B(\frac{ x_3}{|x|}  \pm 1\B )  \frac {-x_1} {x_1^2 + x_2^2 }  \\
= & n x_2  \frac{|x|^2 - x^2_3}{|x|^3}   \frac {-x_1} {x_1^2 + x_2^2 }  \\
= & -  n \frac{x_1 x_2 }{|x|^3}.   \\
\end{split} 
\ee
Thus the second and third  terms in (\ref{stun}) exactly cancel  and hence the result.  
\bigskip

 Now  since $[ (x  \times  \nabla )_k   ,  n x_k |x|^{-1} ]=0$  and $x\cdot (x \times \nabla)=0$
we have that
\be  \label{key} 
\cL^2  =     \sum_k \cL_k^2 =  \sum_k (x \times -i  \nabla )^2_k    +n^2.
\ee
The gauge field has no radial component so  $x \cdot  \nabla   = x \cdot \pa $
and $[ (x \times -i\nabla)_k, |x|^{-1} ]= 0 $
so (\ref{sync2}) becomes
\be \label{H2} 
( f,H f )  = \|  \frac{1}{|x|}(x \cdot   \pa   ) f \|^2   +  (f, \frac{1}{|x|^2} (\cL^2 - n^2 )f ). 
\ee
\bigskip 

Next  change to spherical coordinates. The vector bundle $\pi: E \to M$ becomes 
a vector bundle  $\pi: E' \to \bbR^+   \times S^2$.  With  $U'_{\pm}  \subset   S^2$ defined as in (\ref{four}) 
 these  have  trivializations $h_{\pm}: \pi^{-1} ( \bbR^+ \times U'_{\pm}  )  \to (\bbR^+   \times U'_{\pm} ) \times \bbC$   
 which still have transition functions  $h_+ h_-^{-1} = e^{2in \phi} $. 
The    $ |x|^{-1}  (x \cdot  \nabla  f ) $ becomes $\pa f/\pa r$ and the $\cL_k$ become operators on $\Ga(E') $
specified by  saying that for $x \in \bbR^+ \times U' _{\pm} $,   $(\cL_k  f ) (x) $  satisfies $h_{\pm} (\cL_k f ) (x)  = (x, \cL_k^{\pm} f_{\pm}(x) ) $
where  
  \be
\begin{split}
\cL^{\pm} _1  =&   i  \B( \sin \phi  \frac{\pa}{ \pa \theta } + \cot \theta \cos \phi \frac{\pa}{ \pa \phi  } \B)  - n \cos \phi \left( \frac{1 \mp  \cos \theta}{\sin \theta}  \right)  \\
\cL^{\pm} _2  =&   i  \B( - \cos \phi  \frac{\pa}{ \pa \theta } + \cot \theta \sin  \phi \frac{\pa}{ \pa \phi  } \B)  - n \sin \phi \left( \frac{1 \mp  \cos \theta}{\sin \theta}  \right)  \\
\cL^{\pm} _3 = &  - i  \frac{\pa}{ \pa \phi  } \mp n.  \\
\end{split}
\ee
Note that since  $ ( \pa / \pa \phi) e^{2i n\phi} = e^{2i n\phi } \pa / \pa \phi + 2i n $
we have  in $U'_+ \cap U'_-$  the required  $ \cL^{+ }_i  e^{2i n\phi} =  e^{2i n\phi } \cL^{-}_i $.

The Hamiltonian  in spherical coordinates, still called $H$, has become 
\be  \label{sunny}
( f,H f )  = \|  \frac{\pa  f}{\pa r}  \|^2   +  ( f,  \frac{1}{r^2}  (\cL^2 - n^2)   f ) 
\ee
where now the norms and inner products  are in $\cH  = L^2(E' , r^2  dr   d \Om  )$. 
After an integration by parts this implies
\be \label{sunny2} 
H f=  - \frac{1}{r^2} \frac{\pa} {\pa r} r^2  \frac{\pa f} {\pa r} +  \frac{1}{r^2} ( \cL^2 - n^2).  
\ee
In fact since the
the transition functions only depend on the angular variables we can make the identification 
\be 
\cH =     L^2(\bbR^+ , r^2 dr )  \otimes L^2(\tilde E, d \Om )  
\ee
where $\tilde E$ is a vector bundle $\pi: \tilde E \to  S^2$   with trivializations $h_{\pm}: \pi^{-1}( U'_{\pm} )  \to U'_{\pm}  \times \bbC$   which still satisfy $h_+ h_-^{-1} = e^{2in \phi} $.
Now in  (\ref{sunny2})   the $\cL^2- n^2$ only acts on the factor  $L^2(\tilde E, d \Om)$.

The  joint spectrum of $\cL^2, \cL_3$ has been studied by Wu and Yang \cite{WuYa76}.  
The commutation relations again constrain the possible eigenvalues to $\ell(\ell+1) $ and $|m| \leq \ell$. 
But now from  (\ref{key}) we have $\cL^2 \geq n^2$  so  we must have $\ell \geq |n|$.  Only states with 
non-zero angular momentum exist on the monopole.  
Wu - Yang   explicitly construct the eigenfunctions in term of Jacobi polynomials. 
 The  normalized eigenfunctions    $\cY_{n,  \ell, m} (\theta, \phi)$  are sections of  $L^2(\tilde  E)$
  called \textit{monopole harmonics}.  They satisfy  
\be
\begin{split} 
\cL^2 \cY_{n,\ell,m}= &  \ell(\ell+1)   \cY_{n, \ell,m}   \hs   \ell \geq |n|         \\
 \cL_3  \cY_{n,\ell,m}  = &  m    \cY_{n,\ell,m}  \hs  \hs    |m| \leq \ell.       \\
  \end{split} 
 \ee
 Explicitly they   are given  in  the trivializations on  $U'_{\pm} = U_{\pm} \cap S^2 $ by 
 \be \label{elf1} 
 \cY^{\pm} _{n, l,m}(\xi , \phi )   =  \const (1- \xi)^{\frac12 \al    } (1+ \xi  )^{\frac12 \beta} P^{\al  ,\beta}_{ \ell + m } (\xi )  e^{i(m \pm n) \phi }   \hs \xi = \cos \theta 
 \ee
 where   $\al =  -n-m, \beta = n-m $   and $P^{\al  ,\beta}_{ \ell + m }$ are Jacobi polynomials given    by 
\be \label{elf2} 
P^{\al  ,\beta}_{ \ell + m } (\xi) =  \const (1- \xi)^{- \al    } (1+ \xi )^{ - \beta} \frac{d^{\ell + m}} { d \xi^{\ell + m} }(1- \xi)^{ \al  + \ell +m    } (1+ \xi )^{  \beta + \ell + m}.
\ee
 Completeness follows from the completeness of the Jacobi polynomials. 
Thus the $\cY_{n,\ell,m}$  give the full spectrum of $\cL^2, \cL_3 $ and  yield a definition of these as   self-adjoint operators. 

 Let  $\cK _{n ,\ell} $ be the $2\ell+1$ dimensional  eigenspace in  $L^2(\tilde E, d \Om) $  for the   eigenvalue $\ell( \ell+1)$  of $\cL^2$.  Then  $\cK_{n, \ell} $ is  spanned by the $\{  \cY_{n, \ell, m} \}_{ |m| \leq \ell}   $.    
We now   can write the Hilbert space as
\be
 \cH  = \bigoplus_{\ell =|n| }^{\infty} L^2(\bbR^+, r^2dr) \otimes \cK_{n , \ell} 
 \ee
 and  on smooth functions in this space    $H = \bigoplus_{\ell}\ ( h_{ \ell} \otimes I ) $  where 
 \be    \label{thirsty} 
h_{ \ell}  =   - \frac{d^2}{dr^2} - \frac{2}{r} \frac{d}{dr}   +   \frac{\ell(\ell+1)-n^2 }{r^2}.   
\ee

 The operators $h_{\ell} $ are essentially self-adjoint on $\cC^{\infty} _0( \bbR^+ ) $.  For an operator of this form the   condition is that the coefficient of 
 the $1/r^2 $ term be $\geq \frac34$. (See Reed-Simon \cite{ReSi75}, p. 159- 161 and earlier references).   Here we have $ \ell(\ell+1)  -n^2 \geq \ell \geq |n|  \geq 1$ which suffices.  This means 
 the repulsion from the $1/r^2 $ potential is strong enough to keep the particle away from the origin  and a boundary condition at the origin is not
 needed.  
 
 (This is not the case for the free  radial Hamiltonian $h_{0, \ell}$  with  $n=0, \ell =0$  in which case the $1/r^2$ is absent  and a boundary condition is needed.
 This case does not occur in this paper where $\ell \geq 1$. )  
 
  The self-adjoint  $h_{\ell }$ determines  a unitary group $e^{-ih_{\ell} t} $. This  generates a unitary group on $\cH$ and we  define $H$ to be the self-adjoint
  generator.   So $e^{-iHt}  = \bigoplus_{\ell}  ( e^{-ih_{\ell} t} \otimes I ) $. 
   
\section{Eigenfunction expansions }  \label{tick}

Domains of self-adjointness for  $h_{0, \ell} $  will be obtained by finding continuum  eigenfunction expansions 
(c.f.  Petry  \cite{Pet80}).  But we start with a more general operator   
\be
h(\mu) =     - \frac{d^2}{dr^2} - \frac{2}{r}  \frac{d}{dr}  +  \frac{ \mu^2- \frac14 }{r^2}  
\ee
with $\mu >0$.    As is well-known the continuum eigenfunctions have the form  $ (kr )^{-\frac12} J_{\mu} (kr) $ where  $J_{\mu} $ is the Bessel function of order $\mu$ regular at the origin
and we have
\be
 h(\mu)\B(  (kr )^{-\frac12} J_{\mu} (kr)  \B )  = k^2  \B (  (kr )^{-\frac12} J_{\mu} (kr)  \B  ).    
\ee
Expansions in the eigenfunctions are given by  Fourier-Bessel transforms and we recall the relevant facts.   (See for example Titchmarsh \cite{Tit48}, where
however results are stated with Lebesgue measure $dr$ rather the $r^2dr$ employed here.)  
The  transform
\be
 \psi^{\#}_{\mu}  (k)= \int_0^{ \infty}      (kr )^{-\frac12} J_{\mu} (kr)    \psi (r) r^2  dr
  \ee
defined initially for say $\psi$ in the dense domain   $\cC^{\infty}_0(\bbR^+)$ 
satisfies 
\be
 \int_0^{ \infty}   | \psi^{\#}_{\mu}  (k) |^2 k^2 dk  = \int_0^{ \infty}     |  \psi (r) |^2 r^2  dr
\ee
and  
extends to a unitary  operator  from
$L^2( \bbR^+, r^2 dr)$ to $L^2( \bbR^+, k^2 dk) $.
It   is its own inverse 
\be
 \psi (r)= \int_0^{ \infty}     (kr )^{-\frac12} J_{\mu} (kr)     \psi^{\#} _{\mu}  (k) k^2  dk. 
  \ee
  Now for $\psi^{\#}  _{\mu}  \in  \cC^{\infty}_0(\bbR^+) $   we have that  $\psi(r) $ is a smooth function  and 
\be \label{tidy1}
( h(\mu) \psi)(r) =     \int_0^{ \infty} (kr )^{-\frac12} J_{\mu} (kr)  (  k^2     \psi^{\#} _{\mu}(k)) k^2  dk. 
\ee 
We use this formula to define  $h(\mu)$ as a self-adjoint operator  with domain
\be \label{tidy2} 
 D( h(\mu))  =  \{ \psi \in  L^2( \bbR^+, r^2 dr):   k^2    \psi^{\#} _{\mu}   (k)  \in  L^2( \bbR^+, k^2 dk)   \}. 
\ee
The formula (\ref{tidy1}) provides the spectral resolution  and  so there is a unitary group 
\be \label{tidy3}
( e^{-ih(\mu)t}   \psi)(r) =    \int_0^{ \infty}  (kr )^{-\frac12} J_{\mu} (kr)   e^{-ik^2t}     \psi^{\#} _{\mu}  (k) k^2  dk. 
\ee 
\bigskip

Now  if $\mu = \ell + \frac12 $ then    $\mu^2- \frac14  = \ell(\ell+1) $ and we have the free operator $h_{0, \ell}$.
 Thus with $ \psi^{\#}   = \psi^{\#} _{\ell + \frac12} $ the operator      $h_{0, \ell}  $ is self adjoint on $\{ \psi: k^2    \psi^{\#}(k)   \in  L^2( \bbR^+, k^2 dk)  \}$. 
The unitary group $e^{-ih_{0,\ell}t} $ generates a unitary group on $\cH_0$ and we define $H_0$ to be the self-adjoint generator.  So 
 $e^{-iH_0t} = \bigoplus_{\ell}  e^{-ih_{0,\ell} t} \otimes I$. 

(Note also that if    $\mu =(  ( \ell + \frac12 )^2 -n^2 )^{\frac12} $ then $\mu^2- \frac14  = \ell(\ell+1) - n^2$ and we have the monopole operator $h_{\ell}$.
We do not use this  representation here; see however \cite{Dim20}).

\section{The free dynamics}  \label{tock} 
 We need  more detailed control over the domain of  operator $h_{0, \ell} $  and the associated   dynamics  $e^{-i h_{0, \ell} }$.    
 The  eigenfunctions  can be written  
 \be 
     \frac{1}{ \sqrt{kr}} J_{\ell + \frac12 }(kr)  = \sqrt{ \frac{2}{\pi} }  j_{\ell} (kr) 
  \ee
where    $ j_{\ell} (x) $ are the  spherical Bessel functions 
 which are given for 
 $x>0$ by 
 \be
 j_{\ell} (x) =   \sqrt{ \frac{\pi}{2x} } J_{\ell+ \frac12} (x) =  (-x)^{\ell}\B( \frac{1}{x} \frac{d}{dx} \B)^{\ell}   \frac {\sin x} {x}. 
 \ee
 They are   entire functions which are bounded for  $x$ real and have the asymptotics
 \be \label{asymp} 
j_{\ell} (x ) =
 \begin{cases}  \cO(x^{\ell} ) & x \to 0 \\  \cO(x^{-1}  ) & x \to  \infty. \\
\end{cases} 
\ee
 Now we have the transform pair with $\psi^{\#} = \psi^{\#}_{\ell + \frac12}$ 
 \be  \label{excel} 
 \psi^{\#}  (k)=  \sqrt{ \frac{2}{\pi} }  \int_0^{ \infty}    j_{\ell} (kr)  \psi (r) r^2  dr 
 \hs       \psi (r)=  \sqrt{ \frac{2}{\pi} }  \int_0^{ \infty}    j_{\ell} (kr)  \psi^{\#}   (k) k^2  dk  
  \ee
which   still define unitary operators.  \bigskip

  \begin{lem} \label{integrable} 
  If     $\psi^{\#} \in \cC^{\infty}_0(\bbR^+ )  $ then  for any $N $
   \be \label{slober} 
\psi (r)   =
 \begin{cases}  \cO(r^{\ell} ) & r \to 0 \\  \cO(r^{-N}  ) & r \to  \infty. \\
\end{cases} 
\ee
Furthermore $\psi$ is infinitely differentiable and the the derivatives satisfy for any $N$ 
  \be \label{slober2} 
\psi^{(m)}  (r)   =
 \begin{cases}  \cO(r^{\ell-m} ) & r \to 0 \\  \cO(r^{-N}  ) & r \to  \infty. \\
\end{cases} 
\ee
 \end{lem} 
  \bigskip
 
  \pr
 In (\ref{excel})  $k$ is bounded above and below and so  $ j_{\ell} (kr) $  has   asymptotics  (\ref{asymp})  in $r$
and hence  $\psi(r) $  satisfies (\ref{slober}) with $N=1$.

To improve the long distance asymptotics 
we  use the identity
\be
x  j_{\ell} (x) = (\ell +2  )  j_{\ell+1}  (x)  + x j' _{\ell+1} (x) 
  \ee
to write (\ref{excel}) as 
 \be
 \begin{split}    \sqrt{ \frac{\pi} {2} }  \psi (r)
 = & r^{-1}  \int_0^{ \infty}  k r  j_{\ell} (kr)(  k  \psi^{\#}   (k) )  dk  \\
 = &  r^{-1}    \int_0^{ \infty}  (\ell +2  )  j_{\ell+1}  (kr)   \B(  k \psi^{\#}   (k) \B)dk 
 +    \int_0^{ \infty}  j' _{\ell+1} (kr)  \B( k^2  \psi^{\#}   (k)  \B)  dk.   \\
\end{split}  
\ee
The integral in the first term has the same form that we started with and  we have an extra $r^{-1} $ in front  so   the term is $\cO( r^{-2}) $ as $r \to \infty$.
After integrating by parts the  second term can be written 
 \be
 \begin{split}   
   \frac{1}{r}   \int_0^{ \infty}    \frac{d}{dk}     j_{\ell+1} (kr)    \B(   k^2 \psi^{\#}   (k)  \B)  dk  
 =  &  -   \frac{1}{r}   \int_0^{ \infty}      j_{\ell+1} (kr)  \B(   \frac{d}{dk}(   k^2 \psi^{\#}   (k) )\B)   dk.      \\
\end{split}  
\ee
Again the integral has the same form that we started with and there is an extra $r^{-1}$ so the term  is $\cO(r^{-2} )$.  
Thus we have proved $  \psi (r) = \cO(r^{-2})$ as $r \to \infty$.  Repeating the argument
gives $   \psi (r) = \cO(r^{-N })$ as $r \to \infty$.  Thus (\ref{slober}) is established.

For the derivative we  use $ d/dr ( j_{\ell} (kr) ) = kr^{-1}     d/dk ( j_{\ell} (kr) )$ and integration by parts to  obtain 
\be
 \begin{split}    \sqrt{ \frac{\pi} {2} } \frac{d}{dr}   \psi (r)
 = &  \int_0^{ \infty}    \frac{k}{r}  \frac{d}{dk}   j_{\ell} (kr)  \psi^{\#}   (k)  k^2 dk  \\
 = & \frac{-1}{r}  \int_0^{ \infty}    j_{\ell} (kr)    \frac{d}{dk}  \B(  k^3\psi^{\#}   (k) \B)  dk.   \\
\end{split}  
\ee
The integral is of the same form as  we have been considering  and so has the asymptotics (\ref{slober}).  
But we have an extra  factor $r^{-1} $ and so (\ref{slober2}) is proved for $m=1$.   Repeating the argument
gives the general case.  This completes the proof.  
\bigskip

 The free dynamics (\ref{tidy3})   is now expressed as   
\be  \label{evolve} 
( e^{-ih_{0,\ell} t}   \psi)(r) =    \sqrt{ \frac{2}{\pi} }   \int_0^{ \infty}  j_{\ell} (kr)  e^{-ik^2t}      \psi^{\#}   (k) k^2  dk.  
\ee

 \begin{lem} \label{infty} 
 Let $\psi^{\#} \in \cC^{\infty}_0(\bbR^+ ) $  and $N>0$.  Then there exists a constant $C$ such that  for $0 <   r \leq 1, |t| \geq 1$
 \be
 |  e^{-ih_{0, \ell} t}  \psi (r) | \leq C r^{\ell} |t| ^{-N}.  
 \ee
 \end{lem} 
 \bigskip 
 
 \pr
 In (\ref{evolve})  $k$ is  bounded, hence    $j_{\ell}(kr) = \cO( r^{\ell} )   $ as $r \to 0$,  and hence 
  $ |  e^{-ih_{0, \ell} t}  \psi (r) | $     is  $\cO( r^{\ell} )$  as $r \to 0$ as in the previous lemma.

 Now in  (\ref{evolve}) we write
   \be
  e^{-ik^2t}  =  \frac{1}{-2ikt } \frac{d}{dk}    e^{-ik^2t} 
 \ee
 and then integrate by parts.  This yields
 \be
 \begin{split}
  \label{evolve2} 
& \sqrt{ \frac{\pi}{2} } ( e^{-ih_{0, \ell} t}  \psi ) (r) \\=  &  \frac{1}{2it}   \int_0^{\infty} e^{-ik^2t}  
 \frac{d}{dk}  \B(      j_{\ell} (kr) \  k  \psi^{\#} (k)  \B)  dk   \\
 =  &  \frac{1}{2it}   \int_0^{\infty} e^{-ik^2t}  
 \B( k r     j'_{\ell} (kr)      \psi^{\#} (k)  +       j_{\ell} (kr)  \frac{d}{dk} (k  \psi^{\#} (k) )  \B)  dk   \\ 
 = &  \frac{1}{2it}   \int_0^{\infty} e^{-ik^2t}  
 \B(- k r     j_{\ell+1} (kr)    \psi^{\#} (k)  + \ell  j _{\ell} (kr)     \psi^{\#} (k)   +    j_{\ell} (kr)  \frac{d}{dk} (k  \psi^{\#} (k)   \B)  dk.   \\ 
  \end{split} 
 \ee
 Here we used the identity  
 \be
x j'_{\ell} (x) =   - x j_{\ell+1}  (x) +  \ell j_{\ell } (x).
 \ee
 In the  integral each term  has  the same form we started with (possibly with an extra factor of $r$)  and so are $\cO(r^{\ell} ) $.   But we  have gained a power of   $t^{-1} $ so this   shows that    $ |  e^{-ih_{0, \ell} t}  \psi (r) | \leq  \cO(r^{\ell} |t| ^{-1} ) $
 Repeating the argument gives   the bound $\cO(r^{\ell} |t| ^{-N} ) $.

 \section{Scattering}
 
 Now we are ready  to consider the scattering of a charged particle off a magnetic monopole.   We use
 a two Hilbert space formalism which has been found  useful elsewhere (see  for example \cite{ReSi79}, p 34;  the idea goes back to Kato \cite{Kat67}).
 Recall that the  monopole Hilbert space is the space of sections 
 \be
 \cH  =   L^2(\bbR^+, r^2dr) \otimes L^2(\tilde E, d \Om) = \bigoplus_{\ell =|n|}^{\infty} L^2(\bbR^+, r^2dr) \otimes \cK_{n , \ell} 
\ee  with dynamics $e^{-i Ht}$.  The asymptotic space is 
 \be
 \cH_0   =   L^2(\bbR^+, r^2dr) \otimes L^2(S^2,d \Om ) = \bigoplus_{\ell =0}^{\infty} L^2(\bbR^+, r^2dr) \otimes \cK_{0 , \ell} 
\ee  
 with dynamics  $e^{-iH_0t} $.    To compare them we need an identification operator $J: \cH_0 \to \cH$. 
 We define $J$ by matching angular momentum eigenstates,   taking account that for the monopole only states with $\ell \geq |n|$ occur.
 Thus we define $J$ as a partial isometry  by specifying
 \be
 J  ( \psi  \otimes Y_{\ell,m}  )  =  \begin{cases}   \psi \otimes \cY_{n,  \ell, m}  &  \hs  \ell \geq |n|  \\
 0  & \hs   0 \leq \ell  < n. \\
 \end{cases}   
 \ee
 The M\o ller wave operators are to be defined  on $\cH_0 $ as 
 \be
 \Om_{\pm} \Psi    = \lim_{t \to \pm \infty}  e^{iHt} J e^{-i H_0t}  \Psi 
 \ee
 if the limit exists.
 They vanish  for $\Psi$ in the subspace of $\cH_0$ with $\ell < |n|$.
 The  issue is whether they exist  for $\Psi$ in 
  \be
 \cH_{0, \geq |n|}   \equiv  \bigoplus_{\ell = |n|} ^{\infty} L^2(\bbR^+ , r^2 dr) \otimes  \cK_{0, \ell}. 
 \ee
 If they exist then we have identified states with specified asymptotic form
 \be
 e^{-iHt}      \Om_{\pm} \Psi     \to  J e^{-iH_0t} \Psi   \ \textrm{ as } t \to \pm \infty.
 \ee
 Only states with angular momentum $\ell ( \ell +1),  \ell \geq 1$ occur in the asymptotics. 
  Then we can define  a scattering operator
 \be
 S = \Om_+^* \Om_-
 \ee
 which maps  $\cH_{0, \geq |n|} $ to  $\cH_{0, \geq |n|} $. 
 
 The main result is:

 \begin{thm} \label{one} 
 The wave operators $\Om_{\pm} $ exist.
 \end{thm}
 \bigskip
 
 \pr  For $\Psi \in  \cH_{0, \geq |n|} $ we have   $ \| e^{iHt} J e^{-i H_0t} \Psi \|  =  \| \Psi \| $ so we can approximate $\Psi$ uniformly in $t$ and   it suffices to prove the limit exists for $\Psi$ in a dense set.   
 In fact it suffices to consider  $\Psi = \psi \otimes Y_{\ell, m} $  with 
 with   $\psi^{\#}  \in  \cC^{\infty}_0(\bbR^+ )$ and $\ell \geq |n| \geq 1$  since finite sums of such vectors are dense.
 Since $e^{-iH_0t} \Psi = e^{-i h_{0,\ell} t} \psi \otimes Y_{\ell, m}  $ and $e^{iHt} J\Psi = e^{ i h_{\ell} t} \psi \otimes \cY_{n,\ell, m}$
  the problem reduces to the existence  in $L^2( \bbR^+, r^2 dr ) $  of
 \be
 \lim_{t \to \pm \infty}  e^{ih_{\ell} t}  e^{-i h_{0, \ell} t}  \psi \hs \psi^{\#} \in  \cC^{\infty}_0(\bbR^+ ).
 \ee
 
 To analyze this we need to know that $ e^{-i h_{0, \ell} t}  \psi  \in D(h_{\ell}) $.  It suffices to show that  $ \{ \psi: \psi^\# \in \cC^{\infty}_0(\bbR^+) \} $
 is in $D(h_{\ell})  $.  By lemma  \ref{integrable} this subspace is contained  in the larger subspace
\be
\cD \equiv   \{  \psi \in  \cC^2(\bbR^+):  \psi  \textrm{  has asymptotics  (\ref{slober2})  for }  m=0,1,2   \} 
 \ee
 so it suffices to show $\cD \subset D(h_{ \ell} ) $.
Note that with these asymptotics the derivatives are still in   $L^2( \bbR^+, r^2 dr)$.  Indeed with $m \leq 2$  the worst behavior as $r \to 0$  is $\cO(r^{-1})$ 
and this is still square integrable with the measure $ r^2 dr$.   Thus $h_{ \ell} $ acting as derivatives is an operator on $\cD$ and  by
integrating by parts it is symmetric.    So we have a symmetric extension of the operator  $h_{ \ell} $ on     $\cC^{\infty}_0 (\bbR^+) $ and the latter  is essentially self-adjoint. Thus $h_{\ell}$
on $\cD$ is also essentially self-adjoint with the same closure.    
   In particular $\cD \subset D(h_{ \ell} ) $.

 Let $\Om_t = e^{ih_{\ell} t}  e^{-i h_{0, \ell} t} $.  
 We now can  compute  
   \be 
   (\Om_{t'} -  \Om_{t} ) \psi  = \int_t^{t'}  \frac{d }{ds}  \Om_s \psi ds = \int_t^{t'}  e^{ih_{\ell} s}  ( h_{\ell} - h_{0, \ell}  )  e^{-i h_{0, \ell} s}    \psi.   
\ee   
But 
  \be
  h_{\ell} - h_{0, \ell} =   \frac{-n^2}{r^2} \equiv   v(r) 
  \ee
  and so
  \be
  \| ( \Om_{t'} -  \Om_{t}) \psi  \| \leq    \int_t^{t'}   \| v  e^{-i h_{0, \ell} s}    \psi  \|   ds. 
\ee   

Now it suffices to show that  the function  $t \to  \| v  e^{-i h_{0, \ell} t}    \psi  \| $ is integrable to obtain a limit.   
We write  $v = v_1 + v_2$  with supports respectively in $(0,1]$ and $[1, \infty)$.  Since   $\psi^{\#} \in \cC^{\infty} _0(\bbR^+) $  lemma \ref{infty} 
says $|(  e^{-i h_{0, \ell} t}  \psi )(r) | \leq  \cO( r^{\ell} |t| ^{- N} ) $ for any $N$  and so 
\be \label{bisquit} 
\begin{split}
 \| v_1   e^{-i h_{0, \ell} t}    \psi  \|^2   = &  \int_0^1 \frac{n^4} {r^4} |(  e^{-i h_{0, \ell} t}  \psi )(r) |^2  r^2 dr 
  \leq       \cO(|t|^{ -2N } ) \int_0^1  r^{2 \ell -2}   dr  
  \leq     \cO(|t|^{ -2N } ) \\
\end{split}
\ee   
 which suffices.

For the $v_2$ term we have
 \be \label{olive1} 
 \begin{split}
  \| v_2   e^{-i h_{0, \ell} t}    \psi  \|_{    L^2( \bbR^+, r^2 dr ) } 
 = &      \| v_2   e^{-i h_{0, \ell} t}    \psi  \otimes   Y_{\ell, m}   \|_{    L^2( \bbR^+, r^2 dr ) \otimes L^2 (S^2, d\Om)  } 
\\
 = &      \| v_2   e^{-i H_{0 } t} \Psi    \|_{    L^2( \bbR^+ \times S^2, r^2 dr d\Om)  } 
\\
 = &      \| v_2   e^{-i H_{0 } t} \Psi    \|_{    L^2( \bbR^3 )  }.
\\
  \end{split}
  \ee 
 In the last   step we have returned to  
   Cartesian coordinates, so now $v_2= v_2(|\bx|) $   and      $  e^{-i H_{0 } t} \Psi   $ is  the  unitary   transform of  our 
   $  e^{-i H_{0 } t} \Psi   $ defined  in spherical coordinates.    We want
   to show that this time evolution  is the usual time evolution defined with the Fourier 
transform.  First with $t=0, $   $\Psi$ has become
$
\Psi(\bx) = \psi(|\bx| ) Y_{\ell, m} (\bx/|\bx| ) 
$
 with Fourier transform  
 \be \label{tingle} 
   \tilde \Psi (\bk ) = (2\pi)^{-\frac32} \int e^{i \bk \cdot \bx}  \psi(|\bx| ) Y_{\ell, m} (\bx/|\bx| ) d\bx.
 \ee
 There is a standard expansion of the complex exponential in spherical functions given by the distribution identity (with $k = |\bk|$) 
 \be
 e^{i \bk \cdot \bx}  = 4 \pi \sum_{\ell=0}^{\infty} \sum_{m = - \ell}^{\ell} i^{\ell}j_{\ell} (k|\bx| ) Y_{\ell,m} (  (\bk / | \bk | ) \overline{Y_{\ell,m} (\bx / | \bx| )}.
\ee 
 Inserting this in (\ref{tingle}) and changing back  to spherical coordinates gives 
  \be \label{tingle2} 
   \tilde \Psi (\bk ) =4\pi i^{\ell}\B(   \int_0^{\infty} j_{\ell} (kr)  \psi (r)  r^2 dr \B)    Y_{\ell,m}  (\bk / | \bk | )
    = (2 \pi)^{\frac32}   i^{\ell}  \psi^{\#} (k)  Y_{\ell,m}    (\bk / | \bk | ). 
 \ee
 Now replace   $\psi$ by $e^{-i h_{0, \ell} t} \psi $.  Then $\psi^\#(k) $ becomes $e^{-ik^2t} \psi^\#(k)$ and so $\tilde \Psi(\bk) $ becomes
 $e^{- i |\bk|^2 t } \tilde \Psi(\bk) $.    Thus 
 \be \label{fourf}
( e^{-i H_0t} \Psi)(\bx) = (2\pi)^{-\frac32} \int e^{- i \bk \cdot \bx}  e^{- i |\bk|^2 t } \tilde \Psi(\bk) d\bk 
\ee
 which is the standard time evolution. 
 
  By lemma \ref{integrable} we have that $ \Psi   \in    L^1( \bbR^3, d\bx )  $ since
 \be
 \| \Psi  \|_1 = \int_{\bbR^3 }  \B| \psi (|\bx|)   | Y_{\ell, m}(\bx/  |\bx|    ) \B | d\bx
 = \int_0^{\infty}  |\psi (r) | r^2 dr  \ \int_{S^2 }   |Y_{\ell, m}(\theta, \phi)|  d\Om < \infty.
 \ee
  Thus   $ \Psi   \in    L^1( \bbR^3, d\bx ) \cap     L^2( \bbR^3, d\bx )    $ and in this case (\ref{fourf})  has  
  the well-known representation
  \be
   ( e^{-iH_0t }  \Psi ) (\bx)  =  (4 \pi i t ) ^{-\frac32}  \int e^{ i|\bx-\by|^2/4t}   \Psi (\by) d \by.
   \ee
This  gives the estimate  $ \|  e^{-iH_0t } \Psi   \|_{\infty}  \leq  \cO( |t|^{-3/2} ) $.

 Now  $v_2(|\bx|) $    is  in  $ L^2( \bbR^3, d\bx )  $   (since  $\int_1^{\infty} r^{-4} r^2 dr  < \infty$ ) and so   
   \be  \label{olive2} 
     \| v_2   e^{-i H_0  t}   \Psi   \|_2  \leq   \| v_2 \|_2 \|  e^{-i H_0  t}   \Psi   \|_{\infty} 
     \leq   \cO( |t|^{-3/2} )
  \ee   
  which gives the  integrability in $t$.  This completes the proof.

\section{Perturbations} 

As an indication of the advantages of the present approach we show that it can  accomodate perturbations.   
Let $V(|x|)$ be a smooth bounded    spherically symmetric   function on $\bbR^3$.  This defines a multiplication operator on sections of 
the vector bundle.   We consider instead of the Hamiltonian $H$
the perturbed Hamiltonian $H+V$.  The corresponding radial Hamiltonian for angular momentum $\ell$  is instead of (\ref{thirsty}) 
\be \label{thirsty2} 
h_{\ell} +V =   - \frac{d^2}{dr^2} - \frac{2}{r} \frac{d}{dr}   +\B[  \frac{\ell(\ell+1)-n^2 }{r^2}   + V(r) \B].
\ee
As a bounded perturbation of $h_{\ell}$ which  is essentially self-adjoint on $\cC^{\infty}_0(\bbR)$ it is itself essentially self-adjoint on the same 
domain.    Then $h_{\ell} +V$ generates a unitary group on $L^2(\bbR^+ , r^2dr)$, hence  a unitary group on the full Hilbert space $\cH$,
and $H+V$ is the generator so 
\be
 e^{-i(H+V) t}  = \bigoplus_{\ell =|n|}^{\infty}   ( e^{-i(h_{\ell} +V) t} \otimes I ).
\ee

\begin{thm}
Let  $V(|x|) $ be a smooth   bounded spherically symmetric function in  $L^2(\bbR^3)$. 
Then the wave operators 
\be
\Om_{\pm} (V)\Psi  = \lim_{t \to \pm \infty}  e^{ i(H+V)t } J e^{-iH_0t} \Psi
\ee
exist.
\end{thm}
\bigskip

\pr
As in the proof of theorem  \ref{one}, 
the  proof reduces to showing that 
$\| (v + V ) e^{-ih_{0,\ell} t}  \psi \|$ is integrable in $t$ for $\psi^{\#} \in \cC^{\infty}_0(\bbR^+)$.  
 We already know this for $v = -n^2/r$  so it suffices to consider $\| V e^{-ih_{0,\ell} t}  \psi \|$.  We split
 $V(r)  = V_1(r)  + V_2(r) $ with supports in $(0,1]$ and $[1, \infty)$.    The term   $\| V_1 e^{-ih_{0,\ell} t}  \psi \|$
 is integrable as in  (\ref{bisquit}), in fact it is easier since $V_1$ is bounded.   For the $V_2$ term we follow the argument (\ref{olive1}) -  (\ref{olive2}) 
and obtain the integrability from the condition  $V_2 \in L^2$.
 This completes the proof. 
\bigskip

\rem  One would like to relax the condition that $V$ be bounded  near the origin.    A key feature in  our method is that  $h_{\ell} +V$ should
be essentially self-adjoint on $\cC^{\infty}_0(\bbR^+ )$ and  referring again to \cite{ReSi75}, this is true if the bracketed expression in (\ref{thirsty2})
is greater than or equal to $\frac34 r^{-2} $ near zero.   This expression is bounded below  by $r^{-2} + V(r) $ so it suffices that
\be
V_1(r) \geq - \frac14 \frac{1}{r^2}. 
\ee
 With this hypothesis the self-adjointness holds.   If we also require that $V_1(r) = \cO(r^{-2}) $ as $r \to 0$, as well as $V_2 \in L^2$, 
then the scattering estimates (\ref{bisquit})  and (\ref{olive2}) still hold and the wave operators exist.

Note that the Coulomb potential $V(r)  = \pm q/r $ satisfies the $V_1$ conditions,  however the condition $V_2 \in L^2$ is violated. 
As in ordinary potential scattering the remedy is to modify the free dynamics (see for example \cite{ReSi79}, p. 169).  With this 
modification the wave operators would exist for the Coulomb potential as well.

\end{document}